\documentclass[aip,jcp,twocolumn,showpacs,nobibnotes,reprint]{revtex4-1}
\usepackage{graphics,graphicx,amsfonts,amsmath,amsbsy,amssymb,color}
\usepackage{verbatim}  
\usepackage{psfrag}  
\usepackage{tabularx}
\usepackage{xmpmulti}
\usepackage{hyperref}
\usepackage{marginnote}
\usepackage{subfigure}
\usepackage{paracol}
\usepackage{xparse}
\usepackage{layouts}
\usepackage{afterpage}
\usepackage{braket}
\usepackage{paralist}
\usepackage{bm}
\usepackage{url}

\newcommand{\reffig}[1]{{Fig.~\ref{#1}}}

\newcommand{\nocc}{\ensuremath{n_{\text{occ}}}}
\newcommand{\nvirt}{\ensuremath{n_{\text{virt}}}}

\hypersetup{hyperindex,breaklinks,colorlinks=true,linkcolor=blue,citecolor=blue,urlcolor=blue,filecolor=blue}

\begin{document}

\title{Convergence of many-body wavefunction expansions using a plane wave basis in the thermodynamic limit }

\author{James~J.~Shepherd}
\email{jshep@mit.edu}
\address{Department of Chemistry, Massachusetts Institute of Technology, 77 Massachusetts Ave, Cambridge MA, 02139}

\pacs{71.10.Ca, 71.15.Ap}
\begin{abstract}
Basis set incompleteness error and finite size error can manifest concurrently in systems for which the two effects are phenomenologically well-separated in length scale. 
When this is true, we need not necessarily remove the two sources of error simultaneously.
Instead, the errors can be found and remedied in different parts of the basis set.
This would be of great benefit to a method such as coupled cluster theory since the combined cost of $\nocc^6 \nvirt^4$ could be separated into $\nocc^6$ and $\nvirt^4$ costs with smaller prefactors. 
In this Communication, we present analysis on a data set due to Baardsen and coworkers, containing coupled cluster doubles energies for the 2DEG for $r_s=$ 0.5, 1.0 and 2.0 a.u.~at a wide range of basis set sizes and particle numbers.
In obtaining complete basis set limit thermodynamic limit results, we find that within a small and removable error the above assertion is correct for this simple system. 
This approach allows for the combination of methods which separately address finite size effects and basis set incompleteness error.
\end{abstract}
\date{\today}
\maketitle

{\bf \emph{Introduction}}.--
Since it is extremely challenging to devise methods to simulate an infinite solid directly, a common approach to addressing solid state problems is to use a supercell with a finite particle number and a judicious choice of boundary conditions.
The error made in such an approach is termed finite size error, and represents a substantial road block in the development of realistic wavefunction descriptions of solids.
This is because the error is both substantial and slowly-decaying; for the total energy it commonly falls away as the inverse of the system size, $1/N$.

\begin{figure*}
\begin{center}
\vspace{-1cm}

\makebox[0.8\textwidth][l]{%
\subfigure[\mbox{}]{%
\includegraphics[width=0.4\textwidth]{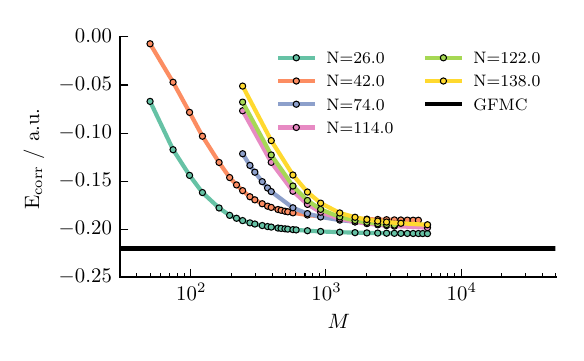}\label{1a}
}
\subfigure[\mbox{}]{%
\includegraphics[width=0.4\textwidth]{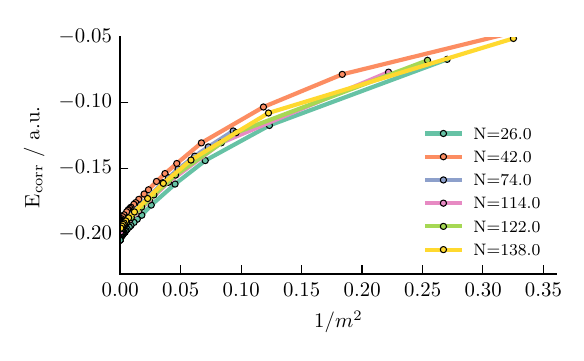}\label{1b}
}
}

\makebox[0.8\textwidth][l]{%
\subfigure[\mbox{}]{%
\includegraphics[width=0.4\textwidth]{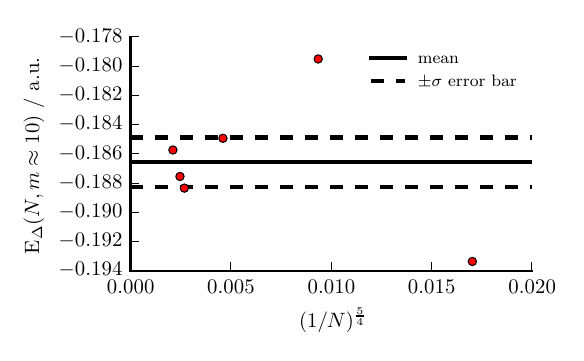}\label{1c}
}
\subfigure[\mbox{}]{%
\includegraphics[width=0.4\textwidth]{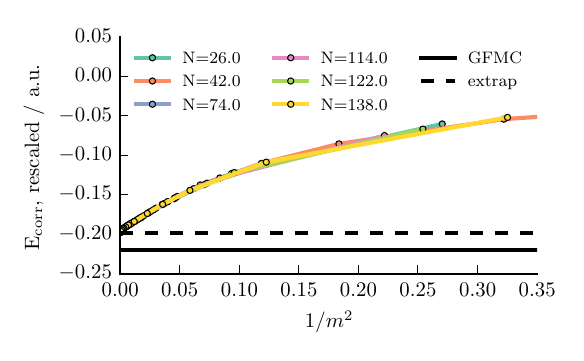}\label{1d} 
}%
}

\caption{The complete basis set limit, thermodynamic limit, coupled cluster doubles correlation energy is obtained for a 2D electron gas at $r_s=1.0$ for comparison with Green's function Monte Carlo. In (a), the raw data (originally from 
Refs.~\onlinecite{Baardsen1} and~\onlinecite{Baardsen2}%
) 
are shown plotted against basis set size $M$ for various particle numbers $N$. In (b), an axis transformation $M\rightarrow1/m^2$ with $m=M/N$ has been performed. It can be seen the the basis set convergence is actually consistent with particle number $N$. In (c), the fluctuation in particle number due to finite size effects are shown to have relatively small variation when de-coupled from basis set energy. In (d), the average value of (c) is added to each line to re-scale each set of points. Now each $N$ value yields a consistent estimate of the final thermodynamic limit, complete basis set limit energy. The final value is $-0.198$ ($\pm 0.004$) Ha.\vspace{-0.5cm}}

\end{center}
\end{figure*}

On account of finite size effects being such a large source of error, many methods have been developed which alleviate them. A review is well beyond the scope of this work (however, see \emph{e.g.} Ref.~\onlinecite{muller_incrementally_2013}), but these methods include the hierarchical~\cite{manby_extension_2006,nolan_calculation_2009} and incremental schemes~\cite{friedrich_incremental_2013,muller_incrementally_2013,fertitta_towards_2016}, progressive downsampling~\cite{ohnishi_hybrid_2011}, and embedding theories for density functionals~\cite{libisch_embedded_2014,goodpaster_exact_2010,manby_simple_2012,goodpaster_density_2012,barnes_accurate_2013} and density matrices~\cite{bulik_electron_2014,bulik_density_2014,knizia_density_2013,knizia_density_2012}. 
Local orbitals~\cite{masur_efficient_2013,usvyat_approaching_2011,maschio_fast_2007,pisani_periodic_2008}, local interactions~\cite{ayala_extrapolating_1999,spencer_efficient_2008, fraser_finite-size_1996,drummond_finite-size_2008,kent_finite-size_1999},  and length scale (including range separation) schemes~\cite{bruneval_range-separated_2012,toulouse_adiabatic-connection_2009,spencer_efficient_2008} have also been developed that exploit a length scale separation between correlations within a unit cell and correlations between unit cells.
Finally, some many-body methods can be directly integrated to find the thermodynamic limit~\cite{bishop_electron_1982,bishop_electron_1978,bishop_overview_1991,nozieres_correlation_1958,gell-mann_correlation_1957,ziesche_self-energy_2007,onsager_integrals_1966} from which analytic corrections can be derived~\cite{chiesa_finite-size_2006}.

Even for a finite system, a finite number of basis functions yields a different type of finite size effect in wavefunction calculations.
A finite set of smooth functions is unable to correctly describe interelectron cusps, and the resultant errors in the energy also have slow convergence in the basis set size. 
In a recent study, convergence of the many-body wave function expansion using a plane wave basis was analyzed for the electron gas and lithium hydride solid~\cite{shepherd_convergence_2012} drawing inspiration from a large body of literature on this subject for molecular systems~\cite{hattig_explicitly_2012}.
When analyzed, the basis set incompleteness error fell off as the inverse of the number of basis functions used: $1/M$. 
The power-laws derived were used in the years leading up to that study and then subsequently to achieve complete basis set results for a variety of systems~\cite{marsman_second-order_2009,gruneis_second-order_2010,booth_towards_2013} and in particular the uniform electron gas~\cite{roggero_quantum_2013,shepherd_range-separated_2014,shepherd_coupled_2014,shepherd_many-body_2013,shepherd_investigation_2012,shepherd_full_2012,spencer_developments_2016,filinov_fermionic_2015,schoof_towards_2015,schoof_textitab_2015,malone_accurate_2016,malone_interaction_2015}.
In recent times, several further major developments addressing plane wave basis set incompleteness error have been made. 
In particular, explicit correlation has been applied to a plane-wave basis, including F12 methods~\cite{gruneis_explicitly_2013,gruneis_efficient_2015,usvyat_linear-scaling_2013} and transcorrelation~\cite{umezawa_ground-state_2004}; corrections have been derived for a semi-analytical correction has been found for the direct term MP2 and for dRPA~\cite{klimes_predictive_2014}; and hybrid basis sets of plane-wave derived occupied orbitals and Gaussian virtual orbitals have been implemented~\cite{booth_plane_2016}.

Converging finite basis sets and finite particle numbers simultaneously to their respective limits is challenging, and a $1/(MN)$ scaling of the error makes even simple methods such as coupled cluster doubles prohibitively expensive in computational cost (scaling as  $\mathcal{O} [ N^6 M^4 ]$ see Ref.~\onlinecite{ohnishi_hybrid_2011}).
A method to alleviate this cost is to undertake calculations with small particle numbers for large basis sets and vice versa, and combining the two results to estimate the result of taking both limits.
Instead of converging the coupled error brute force (at $\mathcal{O} [ N^6 M^4 ]$ cost) this would mean that the two can be removed separately at $\mathcal{O} [ N^6 ] +\mathcal{O} [ M^4 ]$ cost at the penalty of a small, controllable, and analyzable error. 
It is difficult to track the origin of such approaches, with perhaps the earliest mention coming from Nozi{\`e}res and Pines~\cite{nozieres_correlation_1958}; recent authors give attribution to work due to Hirata~\cite{hirata_bridging_2010,hirata_thermodynamic_2012,yamada_asymptotic_2013,hirata_second-order_2014,keceli_fast_2010,shimazaki_brillouin-zone_2009,hirata_fast_2009,ohnishi_hybrid_2011,ohnishi_logarithm_2010}, Kresse~\cite{gruneis_natural_2011}, and colleagues.
Similar physics can also be found in neighboring fields, such as during the construction of Jastrow functions~\cite{ceperley_ground_1978}, frequency theshholding and renormalization~\cite{wen_quantum_2010}, or removal of finite size effects using DFT corrections~\cite{kwee_finite-size_2008}.

The aim of this Communication is to analyze the finite size effects present in the basis set incompleteness error for coupled cluster doubles correlation energies (CCD).
To this end, we perform numerical analysis on the convergence properties of a large data set due to Baardsen and coworkers~\cite{Baardsen1,Baardsen2} which contains CCD energies for the 2D electron gas for $r_s=0.5$, 1.0, 2.0 at a wide range of basis set sizes (up to $M=500$) and particle numbers ($26 \leq N \leq 138$). 
By analyzing the convergence of the basis set incompleteness error as the particle number changes, we find that for these systems basis set incompleteness error and finite size effects are effectively decoupled in the energy and can be removed independently from one another.
There are physical reasons to believe these two limits would not be strongly coupled, however, since they arise from two different limits of the interelectron interaction ($1/r_{12}$).
Basis set incompleteness error arises from the difficulty in describing electron coallescence points, i.e. interactions as $r_{12} \rightarrow 0$, whereas finite size error arises from the long-range interactions of electrons which are being improperly truncated (or mirrored) by periodic supercell approaches.
This being the case, this overall error is strongly system dependent and requires that the physics of the system naturally separates these two effects in range. 
This allows for the benchmarking of CCD against Green's function Monte Carlo methods which are in the complete basis set and thermodynamic limits.

The strategy presented here works directly with the pure plane wave basis suitable for infinite systems. 
This includes the electron gas, but can also be applied to other infinite systems of interest to condensed matter theorists and nuclear physicists. 
In these systems, finite size effects are captured by an electron number, and extreme pathologies can arise from shell-filling finite size errors. 
We rely particularly on work achieved in this area by Drummond and coworkers~\cite{drummond_finite-size_2008}.
It is anticipated that these additional effects will arise in the treatment of complex materials, and we hope that the developments presented here prove of interest to this community as well. 

{\bf \emph{Deriving consistent basis sets with variation in particle number}}.-- 
Coupled cluster doubles (CCD) energies are presented in \reffig{1a} for a 2D electron gas at $r_s=1.0$; these derive from Ref.~\onlinecite{Baardsen1,Baardsen2}. 
We wish to compare the number with the more exact Green's function Monte Carlo (GFMC) result (also shown) which is already both at the complete basis set limit and the thermodynamic limit.
The data are in neither limit and show signs that this might be difficult to achieve. 
Concretely, this is because convergence to the complete basis set limit seems to be at different speeds depending on $N$, and once at the complete basis set limit the value of this also changes with $N$. 
In other words, the basis set incompleteness error and the finite size effects seem inextricably coupled.

This effect arises because, when varying particle number $N$, the basis set index $M$ changes in energy due to changes in the Fermi energy. 
A more consistent measure for basis set size is one which does not change in energy, which, in general, is a function of the number of basis functions per electron $m=M/N$; this is in common with papers which use the energy itself to converge the basis set incompleteness error, but, as we examined in a previous paper, the $M$ gives better convergence properties for the systems shown here~\cite{shepherd_convergence_2012}.
It is also possible to derive relationships for how basis set incompleteness error behaves on approach to the complete basis set limit~\cite{shepherd_convergence_2012,marsman_second-order_2009,gruneis_second-order_2010,klimes_predictive_2014}, which for 2D is $1/m^2$.
When the correlation energies in our data set are plotted in this way, in \reffig{1b}, we can see that the convergence to the CBS is parallel in $m$ for different particle numbers.

{\bf \emph{Linear interpolation for particle-number fluctuations}}.--  By transforming the data set to $(1/m)^2$, energies converge with basis set in a manner that is consistent and invariant with $N$.
Looking at \reffig{1b}, we can see the lines are still offset from one another along the $y$-axis due to finite size effects. 
The trend is non-monotonic due to the data coming from $\Gamma$-point calculations.

We can measure this $N$-dependent shift in the energy. 
Observing that in \reffig{1b} that the lines no longer share $1/m^2$-dependent points, this can be measured by linear interpolation.
For a specific target value of $1/m^2$, we can find the discrete points on either side $(1/m_1^2,E_1)$ and $(1/m_2^2,E_2)$ and then the shift value can be computed as:
\begin{equation}
E_\Delta(N,m)=E_2+\frac{1/m^2-1/m_2^2}{1/m_1^2-1/m_2^2} ( E_1-E_2).
\label{eq:linear}
\end{equation}
Since the lines are parallel, it should not matter unduly what shift we ultimately choose. In general, the larger the value then the worse the finite size effects, and the smaller the value the more expensive the calculations required. 

A plot of this shift value is shown in \reffig{1c} for $m=0.0009$.  
This was chosen for convenience, since all the $N$ values have this basis set size within their ranges. 
The energy shift value $E_\Delta(N,m)$ can be seen to fall with $N$, but in a manner that fluctuates substantially and although we expect the overall limiting scaling to be $N^{-\frac{5}{4}}$, these data do not support an extrapolation.
This might be resolved by twist-averaging the data~\cite{lin_twist-averaged_2001}, but for this work we make do with a relatively crude estimate: the mean and associated error, which comes out at $-0.1866 \pm 0.0041$ Ha.
This may seem like a large value to be computing, but recall that the absolute value of the quantity depends on the value of a target $1/m^2$. 
Instead, we should estimate the size of the effect by the variation with $N$ of this quantity which is 0.02 Ha in range, decreasing to within our error bar for higher $N$.
Crucially, we found that this is not qualitatively sensitive to values of $m = 0.001, 0.1, 0.2, 0.25$. 
In other words, it is possible to use far smaller basis sets to obtain these shift estimates.

We are using a data set from elsewhere, so it is beyond the scope of this work to return to do further calculations in $N$. 
That said, this identifies that this data set requires a greater resolution in $N$, which we now know can be estimated at low $m$ and at minimal cost.  
Furthermore, twist-averaging would improve the trend in $N$ of the energy shift value due to alleviating the shell-filling effects we see.

{\bf \emph{Final extrapolation}}.--
We are now in a position to modify our original graph to show extrapolation to the complete basis set and thermodynamic limits.
By shifting the lines by the quantity $E_\Delta(N)-E_\Delta(\infty)$, we can overlay them and show the complete basis set extrapolation to a thermodynamic limit quantity (\reffig{1d}).
Extrapolating to the combined complete basis set and thermodynamic limits, we find:
\begin{equation}
\begin{split}
E =-0.198 +0.057 \left(N/M\right) + 0.690 \left(N/M\right)^2,
\end{split}
\end{equation}
and so our extrapolated energy is 90$\pm$2 \% of the GFMC energy, which is in agreement with similar previous findings for 3D electron gases~\cite{spencer_developments_2016, shepherd_many-body_2013}.
Repeating this procedure gives $E =-0.251 \pm 0.003$ and $E =-0.1394 \pm 0.0007$ for $r_s=0.5$ and 2.0 respectively.

%

\begin{figure}
\begin{center}
\vspace{-0.5cm}

\makebox[0.45\textwidth][l]{%
\subfigure[\mbox{}]{%
\includegraphics[width=0.22\textwidth]{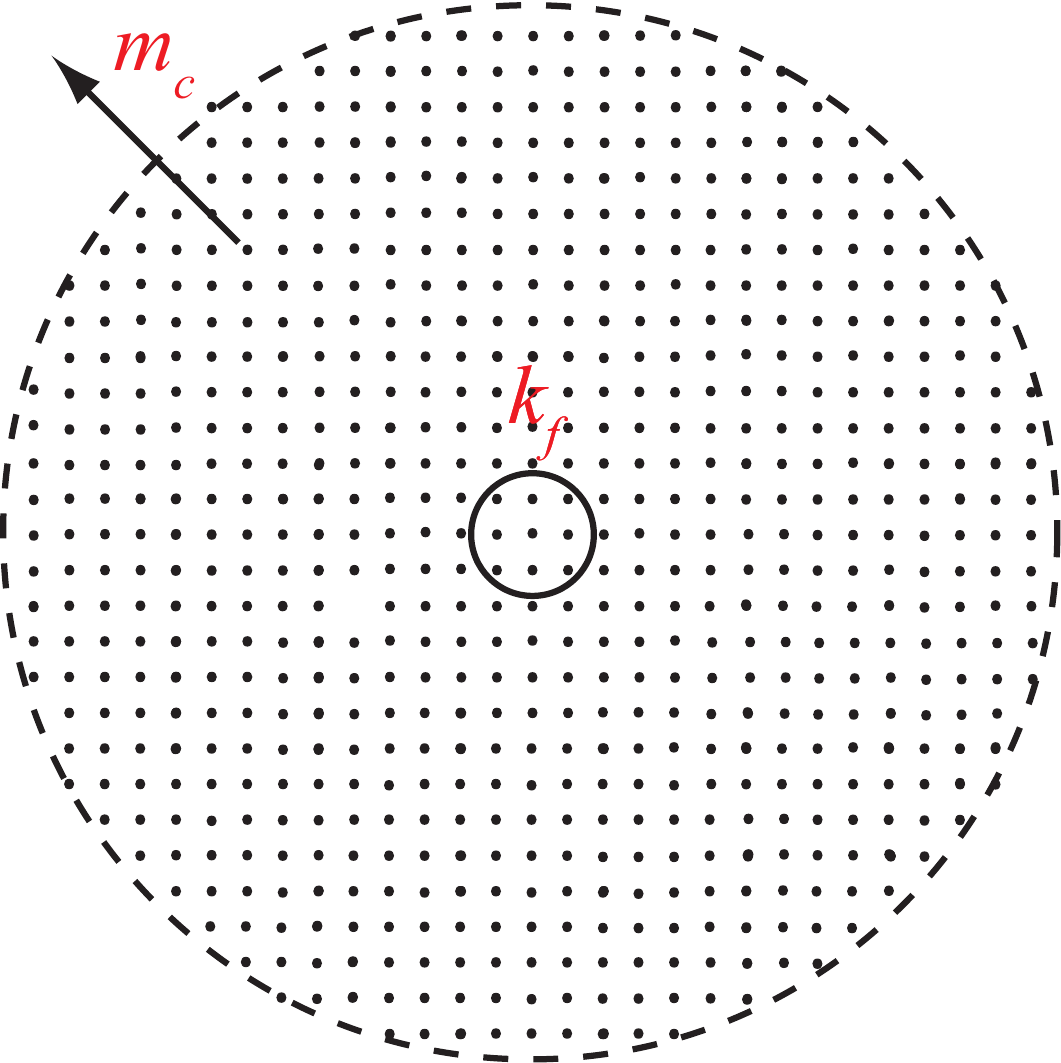}\label{2a}
}
\subfigure[\mbox{}]{%
\includegraphics[width=0.22\textwidth]{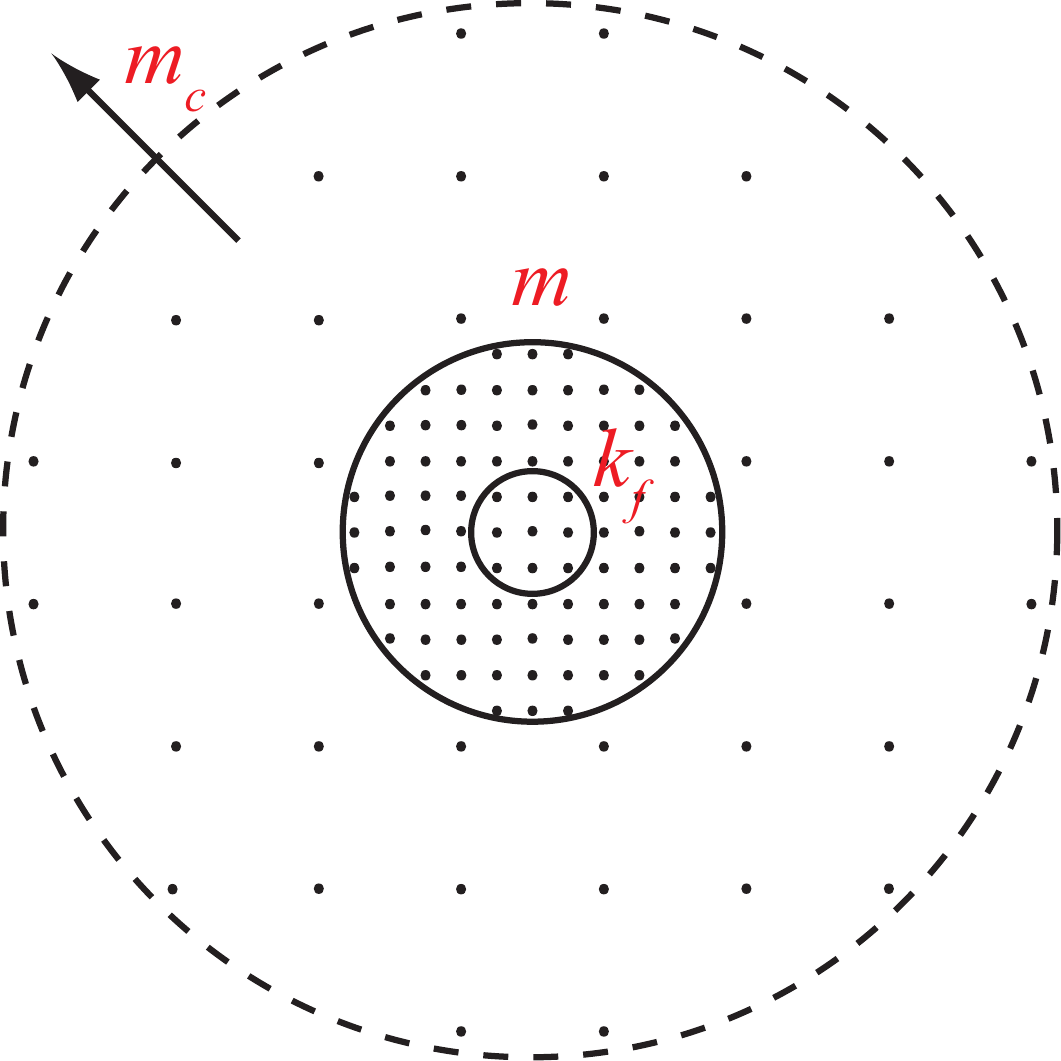}\label{2b}
}
}

\caption{In (a), the conventional strategy is shown where a single particle number (Fermi vector $k_f$) is converged to the complete basis set limit $m_c \rightarrow \infty$. In (b), one value of $m$ is chosen, and the area around the Fermi sphere is gridded finely. Then the basis set limit $m_c\rightarrow\infty$ is found for a small system.}\label{2}
\end{center}
\end{figure}

{\bf \emph{Discussion \& Concluding remarks}}.-- Analyzing Baardsen's coupled cluster doubles data set in the manner presented above reveals that, to a good approximation, the complete basis set and the thermodynamic limits are decoupled. 
The schematic shown in \reffig{2} shows the physical interpretation of such an approach in $k$-space. 
In $k$-space, the thermodynamic limit corresponds to an infinitesimally small grid spacing; the complete basis set limit is reached when an infinite expanse of $k$-space is included.
When we separate both limits, we effectively say we can grid the area around the Fermi surface more finely to converge finite size effects up to a specific basis set limit of $m=M/N$.
In contrast, the area outside of the $m$ cut-off is treated with a more coarse grid and this limit sent out to infinity. 

Provided the finite-$m$ error can be brought under control in the coarse grained part of the space, we can now explore the complete basis set limit at a smaller $N$ and the thermodynamic limit at a smaller $m$ than before. 
This results in substantially improved scaling.

The remaining source of error is two-fold.
The first is that there are still finite size effects in the jagged extrapolation of $N$. 
This could be partially resolved by twist-averaging, and a method for this has been described elsewhere~\cite{shepherd_quantum_2013}.
Further, the data presented here lead us to the expectation that twist-averaging can also be performed at small basis set sizes. 
The second source of error comes from $m$, the inner cutoff, not being large enough/
This causes shell-filling errors and errors caused by coupling between complete basis set limit and the thermodynamic limit.

The finite-$m$ approximation is not as severe as it might look at first glance. 
For sufficiently large basis set sizes (here, $m$) the Fermi sphere becomes point-like compared with the length of the momentum transfer vector which couples the correlated state determinant with the reference. 
We can also argue that there is a limit in which the inter-electron coalescence is not affected by the additional electrons that are provided to the system. 
The additional electrons mediate the fineness of the $k$-point grid, so what this says is that beyond a certain point the fineness of the mesh saturates. 
This is reasonable: the main purpose of a finer grid is to describe the low-lying excitations in the spectrum and to resolve the Fermi sphere. 
In any case, the approximation is less severe as $m$ is raised; in the $m \rightarrow \infty$ limit the expression returns to the original extrapolation scheme without our approximation.%
In order words, we have a controlled approximation with systematic improvability. 
The extent of the coupling between the two regimes and the size of $m$ will depend entirely on physics of the system but is reasonable to think that these two effects are well-separated in length scale~\cite{drummond_finite-size_2008}.

We believe that this provides numerical evidence describing phenomenology that would be of considerable excitement for the community of people examining solid state problems with wavefunction methods. 

{\bf \emph{Acknowledgements.--} } JJS acknowledges Royal Commission for the Exhibition of 1851 for a Research Fellowship.
Thanks is given to Morten Hjorth-Jensen \emph{et. al.} for early access to this data set~\cite{Baardsen1,Baardsen2}.

 \end{document}